# Investigating Crowdsourcing to Generate Distractors for Multiple-Choice Assessments

Travis Scheponik,[1] Enis Golaszewski,[1] Geoffrey Herman,[2] Spencer Offenberger,[2] Linda Oliva,[3] Peter A. H. Peterson,[4] Alan T. Sherman[1]

**Abstract.**
　　We present and analyze results from a pilot study that explores how crowdsourcing can be used in the process of generating distractors (incorrect answer choices) in multiple-choice concept inventories (conceptual tests of understanding). To our knowledge, we are the first to propose and study this approach. Using Amazon Mechanical Turk, we collected approximately 180 open-ended responses to several question stems from the Cybersecurity Concept Inventory of the Cybersecurity Assessment Tools Project and from the Digital Logic Concept Inventory. We generated preliminary distractors by filtering responses, grouping similar responses, selecting the four most frequent groups, and refining a representative distractor for each of these groups.

　　We analyzed our data in two ways. First, we compared the responses and resulting distractors with those from the aforementioned inventories. Second, we obtained feedback from Amazon Mechanical Turk on the resulting new draft test items (including distractors) from additional subjects. Challenges in using crowdsourcing include controlling the selection of subjects and filtering out responses that do not reflect genuine effort. Despite these challenges, our results suggest that crowdsourcing can be a very useful tool in generating effective distractors (attractive to subjects who do not understand the targeted concept). Our results also suggest that this method is faster, easier, and cheaper than is the traditional method of having one or more experts draft distractors, and building on talk-aloud interviews with subjects to uncover their misconceptions. Our results are significant because generating effective distractors is one of the most difficult steps in creating multiple-choice assessments.

[1] Cyber Defense Lab, Department of Computer Science and Electrical Engineering, University of Maryland, Baltimore County (UMBC) - {tschep1, golaszewski, sherman}@umbc.edu
[2] Department of Computer Science, University of Illinois at Urbana-Champaign - {glherman, so10}@illinois.edu
[3] Department of Education, UMBC – oliva@umbc.edu
[4] Department of Computer Science, University of Minnesota Duluth - pahp@d.umn.edu

# 1    Introduction

Generating effective multiple-choice questions (MCQs) is a difficult, time-consuming, and expensive process. For example, in the Cybersecurity Assessment Tools (CATS) Project [She17a], after identifying core concepts and student misconceptions [Par16, Tho18], a group of approximately three to five experts spent several hours per assessment item devising scenarios, question stems, and alternatives (answer choices). A significant part of that effort went into creating the distractors (incorrect alternatives). An effective distractor is attractive to subjects who do not thoroughly understand the targeted concept. We present and analyze results from a pilot study that investigates how crowdsourcing can be used to facilitate the process of generating distractors for concept inventories (conceptual tests of understanding).

Crowdsourcing, for example through *Amazon Mechanical Turk (AMT)* [Kit08], offers several attractive capabilities. One can obtain results from a significant number of diverse respondents inexpensively and with fast turnaround. Stems can be easily pilot tested, and responses provide data about the difficulty of the item and patterns of misconceptions. One could collect such data through in-person interviews, but crowdsourcing facilitates collecting the data more easily, quickly, and cheaply.

We find it especially helpful to ask respondents to enter a free response to a specific stem, without providing any alternative—a technique that can also be used in in-person interviews. If the crowdsourcing responses come from the desired subject population, the responses inherently reflect genuine conceptions and misconceptions. When multiple subjects respond with the same or similar incorrect responses, the response is likely an effective distractor, based on a common misconception. In contrast, in the CATS Project, we identified student misconceptions through interviews before generating some of the stems and all of the distractors. Limited time for interviews prevented us from asking subjects all stems, and we generated or modified some of the stems after analyzing the results of the interviews.

It is necessary to obtain feedback on the development of any assessment item. Generating effective distractors benefits from an iterative process. Crowdsourcing greatly facilitates this iterative process.

One way to use crowdsourcing is to collect and analyze a set of responses to a specific stem. First, a team of experts can filter the collected responses to exclude ones that appear to be unresponsive or lacking genuine thought. Second, the team can group similar responses. Third, the team can select disjoint groups of incorrect responses that appeared frequently. Fourth, the team can refine a representative alternative for each of the selected groups. This process can be useful if it yields a feasible number of quality distractors.

As we explain in Section 2, MCQs in concept inventories should have the following specific properties [McD18]: Each MCQ should target one important concept, there should be exactly one best answer. The alternatives should be homogeneous and disjoint; subjects who understand the concept should find it easy to select the best alternative; and subjects who do not understand the concept should find the distractors attractive [Hes92].

To our knowledge, we are the first to propose using crowdsourcing to facilitate the creation of distractors and the first to conduct a study of this technique. Our hypothesis is that crowdsourcing can be a useful tool for efficiently generating high-quality distractors for multiple choice assessments.

In our experiments, we asked subjects to provide free responses to stems drawn from the *Digital Logic Concept Inventory (DLCI)* [Her10] and the *Cybersecurity Concept Inventory (CCI)* of the CATS Project. Using two different concept inventories helps explore whether crowdsourcing can be useful across diverse domains. In addition, the DLCI provides a well-validated reference against which newly generated distractors can be compared. Validation studies of the CCI are currently underway [Off19].

We analyze subject responses and the resulting distractors in two ways. First, we compare the responses and resulting distractors to those from the DLCI and CCI. Second, also using AMT, we obtain feedback on the resulting new draft test items (including distractors) from additional subjects. Our experiments show that AMT responses from a global set of diverse subjects can be collected and analyzed to generate effective distractors for multiple-choice concept inventories quickly, cheaply, and easily.

In the rest of this paper, we review background and previous work, explain the purpose and methods of our study, present and analyze our results, and discuss issues raised by our study. Our contributions include proposing and investigating the strategy of using crowdsourcing to generate distractors for MCQs in concept inventories.

## 2    Background

We review background material on multiple-choice questions (MCQs), concept inventories, crowdsourcing, the Digital Logic Concept Inventory (DLCI) and the Cybersecurity Concept Inventory (CCI), and the traditional process we used to generate test items in the CCI.

**Multiple-Choice Questions.** Each multiple-choice assessment item comprises a stem (question) and some number (e.g., five) alternatives (answer choices). There is one best alternative; the others are distractors (incorrect choices). Optionally, one or more stems may share a common scenario that provides detailed context, possibly including an artifact such as a figure or table.

**Concept Inventories.** A concept inventory is a multiple-choice criterion-referenced test that assesses student understanding of core concepts and reveals misconceptions (e.g., The Force Concept Inventory [Her10]). They can be used to measure the effectiveness of various teaching approaches.

**Crowdsourcing.** In crowdsourcing, a potentially large number of workers are hired to perform tasks [How06]. Typically, these workers are unaffiliated with the organization requesting the work and they might be anonymous.

**"Cheating" in Crowdsourcing.** A well-known limitation of crowdsourcing is that some subjects do not make a genuine effort to complete the task in good faith. For example, some subjects—including humans and non-humans (bots)—try to earn money by completing surveys as quickly as possible. Subjects who cheat might respond by copying the question, inserting arbitrary text, or leaving the response blank. For example, one of our subjects responded with news about the fashion industry to a question about computer science. The anonymous nature of crowdsourcing motivates and enables some subjects to act unencumbered by social norms [Joi99].

**DLCI.** The Digital Logic Concept Inventory (DLCI) is a concept inventory that has been carefully validated [Her14].

**CCI and CATS Project.** The Cybersecurity Assessment Tools (CATS) Project is developing rigorous instruments that can measure student learning and identify best practices. The first tool is the Cybersecurity Concept Inventory (CCI), which assesses how well students in any first course in cybersecurity understand core concepts.

**Generating MCQs for the CCI.** The CATS team developed assessment items for the CCI using a four-step process. First, we identified core concepts through a Delphi process of cybersecurity experts [Par16]. Second, we conducted talk-aloud interviews to expose student misconceptions [Tho18]. Third, we generated stems and alternatives through in-person and on-line discussions. Fourth, we refined draft questions based on cognitive interviews, expert review, and pilot testing.

## 3  Previous and Related Work

To our knowledge, we are the first to propose using crowdsourcing to generate distractors. For reading comprehension, Araki [Ara16] discusses how natural language processing techniques can be used to generate distractors automatically by extracting events from reading passages. Welbl et al. [Wel7] explore how crowdsourcing can be used to gauge the plausibility of automatically-generated questions. Guo et al. [Guo16] study how automatically-generated questions can enable motivated students to explore additional topics and to reinforce their learning objectives.

Bucholz et al. [Buc11] devise methods for dealing with "cheaters" in their study in which they use AMT to convert text to speech. They developed two metrics for excluding cheaters based on whether the workers were qualified for a task or whether they put forth an honest effort to complete the assigned task. Alonso [Alo08] shows that, by quickly filtering responses for relevance, he can use crowdsourcing to yield high-quality results at low cost.

## 4 Experiment: Purpose and Methods

We explain our experiment, including its purpose, subject population, question stems, participant steps, how we generated distractors using the AMT platform, and how we analyzed the resulting distractors. UMBC's Institutional Review Board approved the protocol.

### 4.1 Purpose

The purpose of our experiment is to investigate the feasibility and desirability of using crowdsourcing to generate distractors for MCQ concept inventories, including assessing the quality of the resulting distractors and the difficulty and cost of generating them.

### 4.2 Subject Population

We invited any 200 AMT workers to complete our survey by posting a "Human Intelligence Task (HIT)" on the AMT site, hoping to obtain at least 50 genuine responses. Subjects were able to find our task from a web user interface that has search capabilities. Seeking greater geographic diversity, we advertised the task in three separate batches, each made available during a separate time period. We have no demographic information about the responding subjects, except that each has an Internet connection and an AMT account. We paid $0.25 to each subject who completed the task, setting a twenty-minute time limit to complete it.

### 4.3 Question Stems

We asked each subject to answer eight question stems (see Appendix A), four from the DLCI (Questions DLCI 1–DLCI 4) and four from the CCI (Questions CCI 1–CCI 4). Figure 1 gives two representative stems, one from each concept inventory. Using stems from these two concept inventories helps establish whether crowdsourcing can be used across diverse disciplines. Also, the DLCI provides a useful reference in that it has been carefully validated [Her14].

**DLCI 1.** A sequential circuit T that has 0 inputs, 3 flip-flops, and 2 outputs. What is the maximum number of distinct states T can potentially be in over time?

**CCI 3**. Alice runs a top-secret government facility where she has hidden a USB stick, with critical information, under a floor tile in her workspace. The facility is secured by guards, 24/7 surveillance, fences, electronically locked doors, sensors, alarms, and windows that cannot be opened. To gain entrance to the facility, all employees must present a cryptographically hardened ID card to guards at a security checkpoint. All of the computer networks in the facility use state-of-the-art computer security practices and are actively monitored. Alice hires Mark (an independent penetration tester) to exfiltrate the data on the USB stick hidden in her workspace.

Choose the strategy that best avoids detection while effectively exfiltrating the data:

*Definitions:*
24/7: Twenty-four hours a day, seven days a week.

**Fig. 1.** Two representative stems, one from the Digital Logic Concept Inventory (DLCI), and one from the Cybersecurity Concept Inventory (CCI).

### 4.4 Methods: Participant Steps

Each subject carried out the following four steps:

1. The subject logged into AMT.
2. The subject entered search criteria to find the survey (hosted via SurveyMonkey).
3. Upon finding the survey in AMT, if the subject accepted the terms of informed consent, they were given access to the remainder of the survey.
4. The participant answered eight free-response questions: four from the DLCI, and four from the CCI.

### 4.5 Methods: Processing Responses

We generated candidate distractors by filtering responses, grouping similar responses, selecting the four largest groups, and refining a representative distractor for each of these four groups.

We filtered out responses that appeared not to reflect a genuine effort to answer the question. Specifically, we filtered out responses that were blank, appeared to answer a different question, responded with a question (perhaps reflecting lack of understanding of what the questions was asking), or copied a question from the survey. We made this determination separately for each response.

To group the remaining AMT responses after filtering each question, we first identified all responses that appeared similar to some original alternative from the DLCI or CCI. Then, we arranged the remaining new alternatives in groups of similar responses.

### 4.6 Methods: Analysis of Distractors

We analyzed the resulting distractors in two ways. First, we compared the free responses from AMT to existing data showing alternatives selected by students in pilot studies of the DLCI and CCI, reporting the results in a series of bar graphs. In this comparison, we report the number of filtered responses, number of responses that overlap existing alternatives from the DLCI and CCI, and number of new alternatives generated from AMT.

Second, we collected data from additional AMT workers who answered the eight questions from the DLCI and CCI, presented in multiple-choice format. We collected these data in two separate surveys: one offered the original alternatives from the DLCI and CCI; the second offered the correct answer and four possibly new distractors from our AMT experiment. We report our results in bar graphs.

As part of the second analysis step, we compared the performance of university students on the DLCI and CCI to the performance of AMT workers on these instruments. We report these results in a stacked bar graph.

## 5 Results

We present results from three surveys that we conducted on AMT in April 2019: Survey 1 collected open-ended responses to eight stems from the DLCI and CCI; Survey 2 administered eight existing test items from the DLCI and CCI to AMT workers; and Survey 3 administered eight modified test items from these concept inventories using distractors generated from Survey 1. After presenting results from these surveys, we compare the performance of AMT workers with that of college and university students. We also give examples of AMT responses to one of the CCI stems.

### 5.1 Survey 1: Generating Distractors from Open-Ended Questions

In Survey 1, without providing any alternatives, we asked AMT workers to answer four open-ended stems from the DLCI and four open-ended stems from the CCI. 539 workers completed the survey, of which 113 (21%) answered in good faith. We discarded 426 (79%) of the responses because they were blank or otherwise nonresponsive.

Figure 2 shows how we categorized the 539 responses. For each of the 113 genuine responses, we classified it as either correct, (incorrect and) overlapping an existing distractor, or (incorrect and) a new distractor.

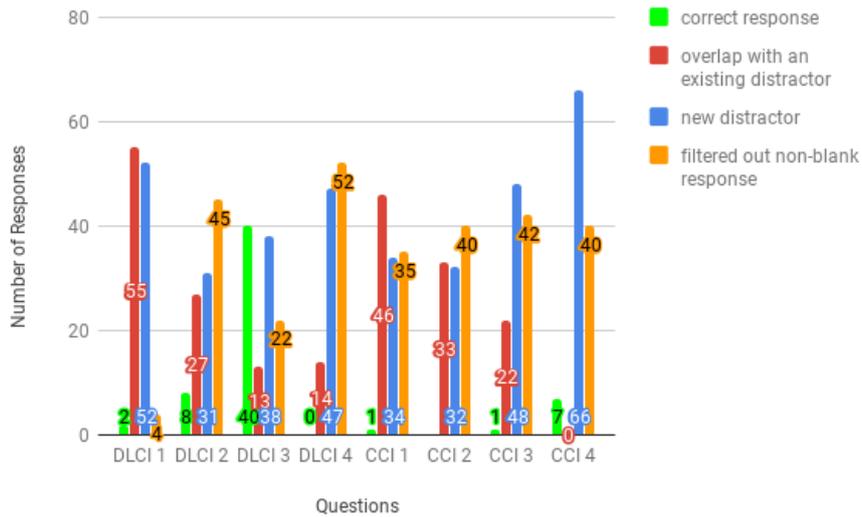

**Fig. 2.** Categorization of 539 AMT responses collected from open-ended stems from the Digital Logic Concept Inventory (DLCI) and Cybersecurity Concept Inventory (CCI). We filtered out 426 responses as blank or otherwise nonresponsive, leaving 113 genuine responses.

**DLCI Distractor Generation**

Questions DLCI 1–4 can be answered with numerals or short sentences.

*Filtering*. DLCI Question 1 had the smallest percentage of filtered responses, perhaps because it asked for a numeric answer.

*Grouping*. Each set of genuine responses resulted in at least four new potential distractors.

*Overlap*. Of the genuine responses, over 50% represent potential new distractors.

*Example*. Question DLCI 2 asked: "Define the word state when used to describe a sequential circuit?" Filtered responses include: "2," "Andhra Pradesh," "tech," "New York," and "Sequential logic is a form of binary circuit design that employs one or more inputs and one or more outputs." Responses that overlap various existing distractors or the correct answer include: "either input output or flip-flop," "State of the circuit depends on the inputs it gets," "memory of element," and "output." The similar responses "If the circuit is on or off," "off-on," "on or off," "whether it is on or off," and others were combined and polished, resulting in the new distractor "State is whether the circuit is 'on' or 'off'."

**CCI Distractor Generation**

Questions CCI 1–4 seek short sentences as answers.

*Filtering.* For each CCI question, we filtered out approximately 40% of the responses.

*Grouping.* Each set of genuine responses resulted in potential new distractors. For each question, the most prevalent two potential new distractors accounted for over 50% of the genuine responses.

*Overlap.* Of the genuine responses, the percent that overlapped existing distractors varied from 0% to 45%.

*Example.* See Section 5.4

## 5.2 Surveys 2-3: AMT Workers Answer Original and Modified Questions from the DLCI and CCI

Figures 3–4 show the performance of AMT workers answering eight original (Fig. 3) and modified (Fig. 4) test items from the DLCI and CCI. The modified questions use the four most popular distractors generated from Survey 1.

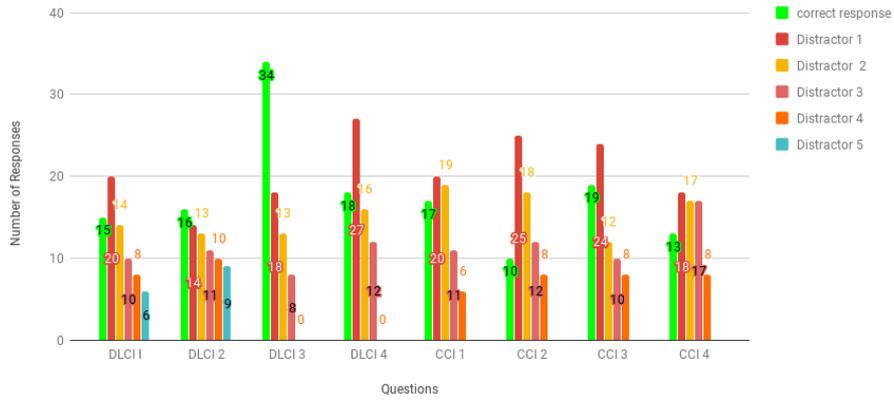

**Fig. 3.** Distribution of 113 responses by AMT workers to eight of the original expert-created multiple-choice questions from the Digital Logic Concept Inventory (DLCI) and Cybersecurity Concept Inventory (CCI). The correct alternative is listed first, followed by the other distractors ordered by their observed frequency. For these test items, AMT workers answered DLCI 3 correctly more frequently than they did for the other questions.

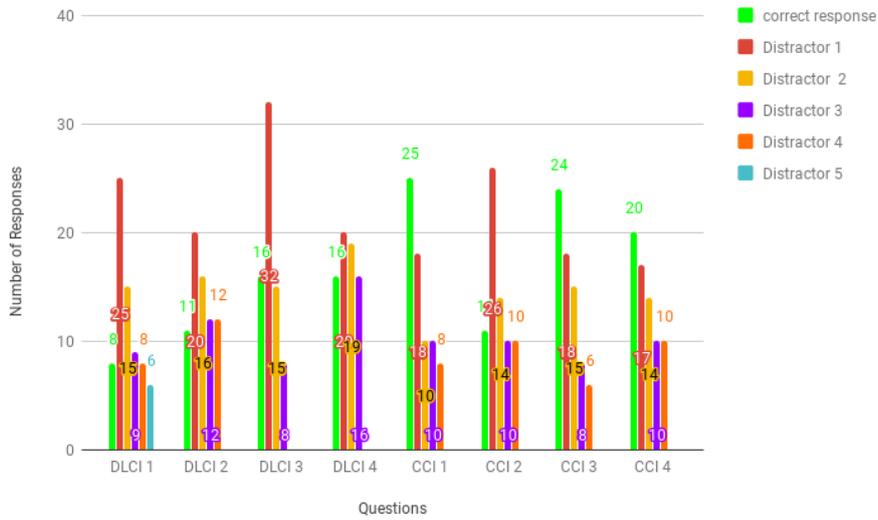

**Fig. 4.** Distribution of 113 responses by AMT workers to eight modified multiple-choice questions from the Digital Logic Concept Inventory (DLCI) and Cybersecurity Concept Inventory (CCI), using distractors generated from Survey 1. For DLCI 3, AMT workers performed worse on the modified question than on the original question (Fig. 3), probably because many of them found the AMT-generated distractor very attractive.

### 5.3 Comparison of Students and AMT Workers on Original Questions from the DLCI and CCI

Using eight original multiple-choice test items from the DLCI and CCI, Figure 5 compares the performance of AMT workers with that of college and university students who participated in validation studies of these concept inventories. These students were taking, or had taken, introductory classes in digital logic or cybersecurity, which studied topics on which the concept inventories are based. On every question, the students outperformed the AMT workers.

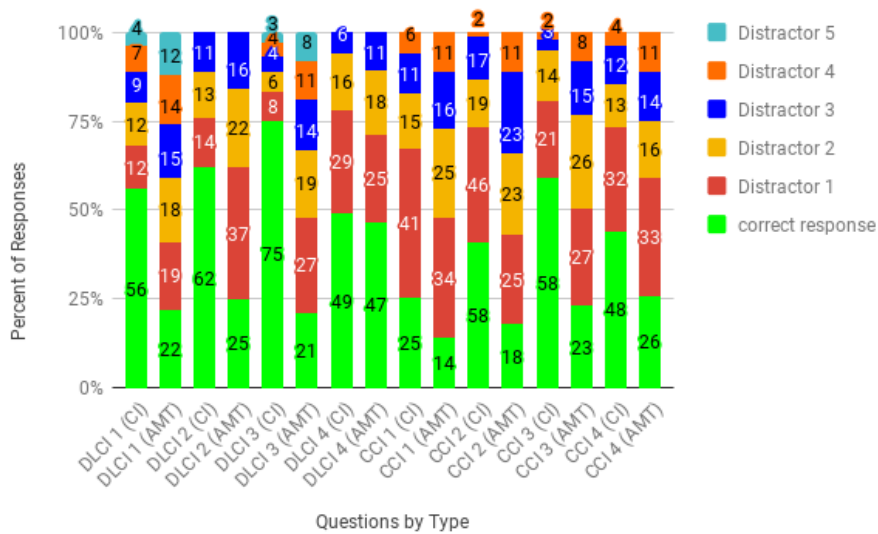

**Fig. 5.** Comparison of student (CI) and AMT worker (AMT) responses to eight original multiple-choice questions from the DLCI and CCI. For every question, students answered correctly more frequently than did AMT workers.

### 5.4 Example of Alternatives Generated from AMT Workers

To illustrate the process and results of using crowdsourcing to generate alternatives, we summarize AMT responses from Question CCI 3 (see Fig. 1).

After a scenario describing an elaborate security system and penetration-testing task, CCI 3 asked, "How can Mark best avoid detection while exfiltrating the data?" First, we filtered AMT responses that did not reflect genuine effort, excluding, for example,

"90," a lengthy quote from a Michelin guide interview with a famous chef, "NO", "yes," "c," "24/8" and many others.

Second, we categorized the remaining responses as correct, overlapping an original distractor, or new. The original alternatives are:

A. Convince an authorized employee to remove the USB stick and give it to Mark. (correct)
B. Compromise the facility's network and add Mark as an authorized guest.
C. Unlock electronically-locked doors using malware.
D. Climb over the perimeter fence at night and sneak into Alice's workspace.
E. Fabricate a fake ID to fool the guards at the security checkpoint.

Examples of responses that overlap various existing alternatives include: "bribe another employee to acquire the item for him" (the correct answer), "Create a copy of an ID card" (overlapping E), and "Get in as guest with current approved employee" (overlapping B).

Third, we grouped the new distractors. For example, we grouped the similar responses, "say no," "to not do it," "avoid for safety," and others.

Fourth, we refined each of the four most popular groups to yield the following refined AMT alternatives:

A. Convince an authorized employee to remove the USB stick and give it to Mark. (correct)
B. Steal an ID card and use this ID to gain access to the facility. (new)
C. Refuse to take the job as security is too strong. (new)
D. Write malicious code to read the data from the USB. (new)
E. Get hired as an employee of the facility. (new)

Filtering and categorizing AMT responses requires human judgment and subject expertise. For example, some AMT responses were somewhat similar to existing distractors but differed in some important details. For example, we feel that stealing an ID is different from fabricating an ID.

As described above, some AMT responses overlapped existing distractors. The AMT responses with new distractors, however, appeared more frequently, so we used them in the test with AMT-derived alternatives.

For this question, AMT workers produced the correct response, responses that overlapped with the original distractors, and new distractors.

## 6 Discussion

We now discuss several issues raised by our pilot study in which, using AMT, we generated distractors for four questions from the DLCI and four questions from the CCI. Specifically, we discuss the quality of distractors generated using AMT, the financial cost and difficulty of generating these distractors, challenges with using crowdsourcing, lessons learned using AMT, limitations of our study, suggestions for improving AMT, and open problems.

### 6.1 Quality of Distractors Generated Using Mechanical Turk

Our pilot study generated many quality distractors, including ones similar to distractors on the DLCI and CCI and new distractors. Subsequent testing of the resulting complete test items on AMT with new subjects confirmed the attractiveness of the distractors to AMT workers. It remains to be determined whether other target audiences will also find these distractors appealing.

### 6.2 Cost and Difficulty of Generating Distractors Using Mechanical Turk

Using AMT, we can easily collect many (e.g., 200) responses overnight, paying $0.25 a worker per task (set of responses). The resulting data can be processed and analyzed in a small number of hours.

In this study, for the CCI, we spent a total of four hours and twenty-five minutes to analyze, filter the responses, and generate new distractors. For the DLCI, we spent a total of one hour and ten minutes. This difference in time results primarily because the DLCI answers are largely numeric, whereas the CCI answers are mostly prose.

In comparison, as explained in Section 2, the traditional process we used to develop the CCI involved many hours of work by a team of experts, carried out over a period of weeks. The process of interviewing subjects and analyzing the resulting transcripts is difficult.

### 6.3 Challenges of Crowdsourcing

It is difficult to control the selection of subjects and to ensure that subjects are human. Motivated in part by financial gain, some may try to "cheat" by plagiarizing responses or by not making a genuine attempt to answer the stem. The potentially large set of responses must be processed. Unlike in-person interviews, crowdsourcing does not permit the interviewer to deviate from the script interactively.

For some questions, AMT workers produced responses that were "too good" to include, given that there is supposed to be one best (not necessarily ideal) alternative. In some of these cases, while developing the CCI, we had previously considered and rejected these responses. Test developers sometimes find it convenient to direct a subject's reasoning about a chosen concept by intentionally excluding certain powerful alternatives.

### 6.4 Lessons Learned

The choice of the worker reward and deadline can influence the likelihood of obtaining the desired data and possibly the quality of results. We offered $0.25 a task with a twenty-minute deadline and found that those choices typically produced at least one hundred responses by the deadline. Currently, a reward of $7 is needed to appear on the first page of advertised tasks, but being on this first page does not seem to be very important.

Responses suggest that many workers were candid, for example stating that they did not understand a question. We speculate that the anonymity of the process encouraged many workers to write statements that some subjects at in-person interviews might feel embarrassed to utter.

We found it useful to direct workers to an on-line survey (using SurveyMonkey) to facilitate collecting the desired information. We did not restrict subjects by answers to demographic questions because we did not have confidence that they would answer truthfully.

Given that AMT is available throughout the world, in an attempt to increase the diversity of the respondents across time zones, we performed multiple batches of work at different times of the day. For example, for each task, we performed three batches each separated by approximately six to eight hours. In different batches, the team tended to receive responses from users in different time zone. For example, students from the University of Hawaii and Eastern Kentucky University responded in different batches. We, however, have not scientifically analyzed the effects of this strategy.

### 6.5   Study Limitations

Limitations of our study using AMT include very limited control of the subject population, the impossibility of verifying whether subjects had the credentials we sought, and the absence of traditional validation studies of the resulting test items. Other limitations include sample size and that the process of grouping the distractors inherently requires some subjective judgement.

### 6.6   Suggestions for Improving Mechanical Turk

It would greatly improve the value of the services offered by AMT if it were possible to control the characteristics of the subjects more reliably. For example, we would have liked to restrict subjects to people who have recently taken a first course in cybersecurity. AMT does offer "premium qualifications," but we would like more choices and greater assurance that subjects have the characteristics that they purport to have. For example, AMT offers qualifications based on income, but this qualification has no bearing on our study. Relatedly, it would be helpful to obtain more information about each worker who completes any task. For example, it would be useful to know the reputation of the worker based on their history and whether the worker is human. In principle, it would be possible to enforce many such conditions by requiring subjects to register and present cryptographically-verifiable credentials, though such credentials are not readily available today. Such information could be used to filter responses.

### 6.7   Open Problems

It would be interesting to explore in greater detail how crowdsourcing can be used to support the entire process of creating a concept inventory, from identifying core concepts through validating draft inventories. It would be helpful to learn more about different methods for generating distractors, the quality of the distractors produced, and the effects of reward pricing. The most robust way to evaluate the quality of distractors is to engage in traditional validation studies of resulting test items, including expert review and large-scale psychometric testing [Her10].

Early in our work we had considered the possibility of evaluating distractors by experts, where the experts would rate distractors not knowing their source. One could thereby compare distractors generated by different methods, including using crowdsourcing and traditional methods. Ultimately, we chose not to carry out such an evaluation because experts are known to be weak arbiters of the attractiveness of distractors [Nat01].

# 7 Conclusion

Our pilot study shows that crowdsourcing can be a tremendously efficient method for generating distractors for multiple-choice concept inventories, saving developers significant amounts of time and money when compared to the more common and much more costly approaches of discussions and in-person interviews.

Using AMT, we generated distractors for the DLCI and CCI, including new distractors and ones similar to those on these concept inventories. The process also produced useful feedback from subjects on stems that they found unclearly worded. Crowdsourcing is very well suited for repeating work tasks with modified material.

Challenges of crowdsourcing include limited control of subjects, the need to filter out responses that do not reflect genuine effort, and the effort required to process the responses and to refine the resulting distractors. We found that filtering and processing the responses was fairly easy to do, and all distractors must be refined regardless of how they were generated. Even though a high percent of our responses were of low quality, the process was still very useful because it generated a sufficient number (at least four) of quality distractors.

Although we focus on concept inventories, we conjecture that crowdsourcing is also useful to develop other types of MCQs. Our promising initial results inspire us to continue to use, study, and refine crowdsourcing to generate distractors. We encourage others to do likewise.

**Acknowledgments**

This work was supported in part by the U.S. Department of Defense under CAE-R grants H98230-15-1-0294, H98230-15-1-0273, H98230-17-1-0349, and H98230-17-1-0347; and by the National Science Foundation under SFS grants 1241576, 1753681, and 1819521, and DGE grant 1820531.

**Appendix A: DLCI and CCI Question Stems**

**A.1 Selected Stems from the Digital Logic Concept Inventory**

**DLCI 1.** A sequential circuit T that has 0 inputs, 3 flip-flops, and 2 outputs. What is the maximum number of distinct states T can potentially be in over time?

**DLCI 2.** Which statement best defines the word state when used to describe a sequential circuit?

**DLCI 3.** A combinational circuit is specified by the truth table below. For three input combinations, the output of the circuit does not matter ("don't-care"). The specification is implemented as a circuit using the following Boolean expression: $f = ac + b$. What will the circuit output when it receives the input combination <a,b,c> = <1,1,0>?

    0 0 0 | 1

    0 0 1 | 0

    0 1 0 | 1

    0 1 1 | 1

    1 0 0 | 0

    1 0 1 | X (don-t care)

    1 1 0 | X (don-t care)

**DLCI 4.** Why do computers use two's complement number representation?

**A.2 Selected Stems from the Cybersecurity Concept Inventory**

**CCI 1.** Bob's manager Alice is traveling abroad to give a sales presentation about an important new product. Bob receives an email with the following message: "Bob, I just arrived and the airline lost my luggage. Would you please send me the technical specifications? Thanks, Alice."

Upon receiving Alice's message, choose what action Bob should take?

**CCI 2.** A company delivers packages to customers using drones. The company's command center controls the drones by exchanging messages with them. The company's command center authenticates each message with a keyed message authentication code (MAC), using a key that is known by the command center and installed in each drone at initialization. The command center stores this key encrypted in a database.

Choose the most promising action for a malicious adversary to masquerade as the command center:

*Definitions:*
to masquerade: To pretend to be someone else.

**CCI 3.** Alice runs a top-secret government facility where she has hidden a USB stick, with critical information, under a floor tile in her workspace. The facility is secured by guards, 24/7 surveillance, fences, electronically locked doors, sensors, alarms, and windows that cannot be opened. To gain entrance to the facility, all employees must present a cryptographically hardened ID card to guards at a security checkpoint. All of the computer networks in the facility use state-of-the-art computer security practices and are actively monitored. Alice hires Mark (an independent penetration tester) to exfiltrate the data on the USB stick hidden in her workspace.

Choose the strategy that best avoids detection while effectively exfiltrating the data:

*Definitions:*
24/7: Twenty-four hours a day, seven days a week.

CCI 4. When a user Mike O'Brien registered a new account for an online shopping site, he was required to provide his username, address, first and last name, and a password. Immediately after Mike submitted his request, you –as the security engineer– observe a database input error message in the logs.

For this error message, choose the potential vulnerability that warrants the most concern:

**Appendix B: Refined AMT Alternatives**

**DLCI 1.**
- **A)** 0 states.
- **B)** 1 state.
- **C)** 2 states.
- **D)** 3 states.
- **E)** 5 states.
- **F)** 8 states.

**DLCI 2.**
- **A)** State is the current value of all storage elements of the circuit.
- **B)** State is the amount of flow (energy) that the circuit can utilize.
- **C)** State is whether the circuit is "On" or "Off".
- **D)** State is the outputs of the circuit.
- **E)** State is the combination of all components of the circuit.

**DLCI 3.**
- **A)** 0
- **B)** 1
- **C)** 2
- **D)** X (don't care)

**DLCI 4.**
- **A)** Computers must use the binary representation of a number.
- **B)** Improves the safety and redundancy of the computer.
- **C)** Communication within the computer is easiest to decode.
- **D)** Subtraction circuits (separate from addition circuits) are not needed in two's complement representation.

**CCI 1.**
- **A)** Send Alice an encrypted email with the technical specifications.
- **B)** Contact the airline so that Alice can find her luggage.
- **C)** Report Alice to internal security for losing the documents.
- **D)** Verify that Alice sent the email requesting the technical specifications.
- **E)** Fly out to Alice's location with the technical specifications.

**CCI 2.**
- **A)** Capture a drone and extract its secret key.
- **B)** Intercept the MAC and replace the MAC with your own.
- **C)** Place a fake order and force the drones to deliver your package.
- **D)** Hack into the database to issue new commands.
- **E)** Jam the drones with your own signals so that the command center cannot communicate with the drones.

**CCI 3.**
- **A)** Convince an authorized employee to remove the USB stick and give it to Mark.
- **B)** Steal an ID card and use this ID to gain access to the facility.
- **C)** Refuse to take the job as security is too strong.
- **D)** Write malicious code to read the data from the USB.
- **E)** Get hired as an employee of the facility.

**CCI 4.**
- **A)** Mike's input may cause the database to execute unintended code.
- **B)** Mike's password was the same as another user's password.
- **C)** Mike's information was not saved in the database.
- **D)** Mike registered his account on a fake website and Mike will be unable to make purchases.
- **E)** Mike was able to provide false information to the database.